\documentstyle[epsfig,12pt,preprint,tighten,aps]{revtex}
\begin{document}

\draft

\title{\rightline{{\tt (October 1998)}}
\ \\
Comment on\\ 
``Big Bang Nucleosynthesis and Active-Sterile Neutrino Mixing:\\
Evidence for Maximal $\nu_{\mu} \leftrightarrow \nu_{\tau}$ Mixing 
in Super Kamiokande?''}
\author{R. Foot and R. R. Volkas}
\address{School of Physics\\
Research Centre for High Energy Physics\\
The University of Melbourne\\
Parkville 3052 Australia\\
(foot@physics.unimelb.edu.au, r.volkas@physics.unimelb.edu.au)}

\maketitle

\begin{abstract} 
The paper ``Big Bang Nucleosynthesis and Active-Sterile Neutrino 
Mixing:
Evidence for Maximal $\nu_{\mu} \leftrightarrow \nu_{\tau}$ 
Mixing in Super Kamiokande?'' by X.
Shi and G. M. Fuller (astro-ph/9810075) discusses the cosmological 
implications of the $\nu_{\mu} \to \nu_s$
solution to the atmospheric neutrino anomaly. It incorrectly 
concludes that a lower bound on the
$\nu_{\tau}$ mass of 15 eV is needed in order for the $\nu_{\mu} \to \nu_s$ 
oscillations to be
consistent with a Big Bang Nucleosynthesis upper bound on the 
effective number of neutrino flavours of
3.3. Since such a large $\nu_{\tau}$ mass is disfavoured from 
large scale structure formation
considerations, the strong, but incorrect, 
conclusion is made that cosmology favours the
$\nu_{\mu} \to \nu_{\tau}$ solution to the atmospheric neutrino 
problem. We explain the nature
of the error. We conclude that cosmology is, at present, consistent 
with both the $\nu_{\mu} \to
\nu_s$ and $\nu_{\mu} \to \nu_{\tau}$ possibilities.

\end{abstract}

\newpage

The large up-down asymmetry observed by SuperKamiokande 
for atmospheric neutrino
induced $\mu$-like events provides compelling evidence for the 
disappearence of muon-neutrinos\cite{sk}.
The most natural explanation of this phenomenon is neutrino mass 
driven oscillations between
$\nu_{\mu}$ and another flavour $\nu_x$, with oscillation 
parameters in the approximate range\cite{fvy}
\begin{equation}
10^{-3} < \Delta m_{\mu x}^2/\text{eV}^2 < 10^{-2},\qquad 
0.8 < \sin^2 2\theta_{\mu x} < 1,
\label{oscpars}
\end{equation}
where $\Delta m^2_{\mu x}$ is the squared mass 
difference, and $\theta_{\mu x}$ is the
vacuum mixing angle. The SuperKamiokande results cannot at 
present distinguish between the
competing $\nu_x = \nu_s$ and $\nu_x = \nu_{\tau}$ possibilities, 
where $\nu_s$ is a sterile neutrino.

The $\nu_s$ solution appears to provide a cosmological challenge: 
with the parameters as per
Eq.(\ref{oscpars}) naive calculations\cite{naive} suggest that the $\nu_s$ is 
thermally equilibrated prior
to the Big Bang Nucleosynthesis (BBN) epoch, leading to an effective 
number of neutrino flavours
$N_{\text{eff}} = 4$. It is at present unclear if $N_{\text{eff}} = 4$ is 
inconsistent, but the low
primordial deuterium abundance result reported in Ref.\cite{low} 
favours $N_{\text{eff}} \ll 4$. The
naive calculations referred to above ignore the 
creation of neutrino-antineutrino asymmetries by
active-sterile neutrino oscillations\cite{ftv}. 
Large neutrino-antineutrino asymmetries suppress
$\nu_{\mu} \to \nu_s$ oscillations, and generically invalidate 
conclusions drawn from the naive calculations.

Neutrino asymmetry evolution as driven by active-sterile 
oscillations goes through two distinct phases at 
temperatures immediately below the critical temperature $T_c$ at 
which neutrino
asymmetry growth begins. The first phase is explosive growth that 
initially is exponential in
character. After a time, the 
non-linear evolution equation for the neutrino asymmetry turns
the exponential growth into power law $T^{-4}$ growth.

We have shown in Ref.\cite{fv} that small angle $\nu_\tau \to \nu_s$ 
oscillations with $m_{\nu_{\tau}} >
m_{\nu_s}$ will, for a range of
parameters, create a large $\nu_\tau$
asymmetry which will then suppress $\nu_{\mu} \to \nu_s$ 
oscillations. 
The situation is a little complicated because
$\nu_\mu \to \nu_s$ oscillations act to create
a $\nu_\mu$ asymmetry which can compensate
for the effect of the large $\nu_\tau$ asymmetry
in the effective potential for $\nu_\mu \to \nu_s$
oscillations. Indeed the Wolfenstein part of
the effective potentials\cite{nr} for $\nu_\tau \to \nu_s$
oscillations and $\nu_\mu \to \nu_s$ oscillations
are respectively proportional to the quantities
\begin{eqnarray}
L^{(\tau)} \sim 2L_{\nu_\tau} + L_{\nu_{\mu}},
\nonumber \\
L^{(\mu)} \sim 2L_{\nu_\mu} + L_{\nu_{\tau}},
\end{eqnarray}
where $L_f \equiv (n_f - n_{\bar{f}})/n_{\gamma}$ is the asymmetry for
species $f$ whose number density is denoted by $n_f$.
The generation of $L_{\nu_\tau}$ asymmetry
by $\nu_\tau \to \nu_s$ oscillations will
also generate a large $L^{(\mu)}$ asymmetry,
which then suppresses $\nu_\mu \to \nu_s$ oscillations
provided that these oscillations do not generate a
significant $L_{\nu_\mu}$ asymmetry in such a way as to drive 
$L^{(\mu)}$ to zero. 
Whether or not the $\nu_\mu \to \nu_s$ oscillations
can compensate for the effects of
the large $\nu_\tau$ asymmetry will depend on
the {\it rates} at which $L_{\nu_\tau}$ and $L_{\nu_\mu}$
are generated.
The rates at which $L_{\nu_\tau}$ and $L_{\nu_{\mu}}$
are generated depend on the oscillation parameters,
$\Delta m^2$ and $\sin^2 2\theta$, for each oscillation mode.
If we fix $\Delta m^2_{\mu s}$ and $\sin^2 2\theta_{\mu s}$ to fit the
atmospheric neutrino anomaly then this leaves two
free parameters, $\Delta m^2_{\tau s}$ and $\sin^2 2\theta_{\tau s}$ for
the $\nu_\tau \to \nu_s$ oscillations.
The issue therefore is to calculate the region
of $(\sin^2 2\theta_{\tau s}, \Delta m^2_{\tau s}$) parameter
space where the large $\nu_\tau \to \nu_s$
oscillation generated $\nu_\tau$ asymmetry
is not compensated by a $\nu_\mu \to \nu_s$
oscillation generated $\nu_\mu$ asymmetry.
For reasons explained in a moment, this issue
is best handled numerically.
The numerical integration of the coupled
evolution equations for $L_{\nu_{\tau}}$ and $L_{\nu_{\mu}}$
for this system was first done in Ref.\cite{fv}, and updated
in Ref.\cite{f}. See Figure 7 of Ref.\cite{fv} and 
Figure 2 of Ref.\cite{f} for the pertinent results.
These results show that $m_{\nu_\tau} > few\ eV$ is 
required, with the precise value being a function of $\sin^2 2\theta_{\tau s}$.
Thus, the interesting conclusion is that if the atmospheric neutrino
anomaly is due to $\nu_\mu \to \nu_s$ oscillations
then the $\nu_{\tau}$ must have a cosmologically interesting
mass. Furthermore, the mass of the $\nu_\tau$ should
be in the range which can be probed by the
Chorus and Nomad experiments.

Also contained in Ref.\cite{fv} (see section VI) was 
an approximate analytic computation
of the allowed region of parameter space. The main difficulty
in solving this problem analytically is that the evolution
equations for $L_{\nu_{\tau}}$ and $L_{\nu_\mu}$ are complicated coupled 
non-linear
integro-differential equations. For the approximate analytic computation of 
Ref.\cite{fv} we made the simplifing
assumption that the $\nu_\tau$ asymmetry is in the $T^{-4}$ 
phase when $\nu_{\mu} \to \nu_s$
oscillations are most strongly acting to produce 
a compensating $L_{\nu_\mu}$.
This simplifing assumption makes things much easier
since it means that
\begin{equation}
{dL_{\nu_\tau} \over dT} \sim {-4L_{\nu_\tau} \over T}.
\label{x1}
\end{equation}
With this simplifying assumption (plus some other assumptions)
we obtained the allowed region of parameter space
where the $\nu_\tau \to \nu_s$
oscillation generated $L_{\nu_\tau}$ asymmetry successfully
suppressed maximal $\nu_\mu \to \nu_s$ oscillations. The
result was the following simple expression [see
Eq.(144) of Ref.\cite{fv}]
\footnote{Note that the meaning of this bound,
assuming for the moment that it is exactly valid, 
is as follows: If it is
obeyed, then negligible numbers of sterile neutrinos are produced by
$\nu_{\mu} \to \nu_s$ oscillations. If it is not obeyed, then
the sterile neutrinos will eventually be brought into thermal
equilibrium by $\nu_{\mu} \to \nu_s$ oscillations.},
\begin{equation}
|\Delta m^2_{\tau s}/\text{eV}^2| \stackrel{>}{\sim}
6 \times 10^5 \left(|\Delta m^2_{\mu
s}/\text{eV}^2|\right)^{12/11}
\label{naivebound}
\end{equation}
This result is independent of $\sin^2 2\theta_{\tau s}$ 
just because Eq.(\ref{x1}) is independent of 
$\sin^2 2\theta_{\tau s}$.
However, our numerical work (also in section VI of Ref.\cite{fv})
showed that the simplifying assumption
that leads to Eq.(\ref{x1}) is not valid. The reason is that
the $\nu_\tau$ asymmetry is still in the
explosive growth phase when $\nu_{\mu} \to \nu_s$
oscillations are most strongly acting to produce a 
compensating $L_{\nu_\mu}$ asymmetry.
This means that the rate at which $L_{\nu_\tau}$ is generated
is actually much greater than assumed in the 
derivation of Eq.(\ref{naivebound}). 
This makes the allowed
region {\it significantly} larger and it also 
depends on $\sin^2 2\theta_{\tau s}$
(since the rate of $L_{\nu_\tau}$ generation depends
on $\sin^2 2\theta_{\tau s}$ in the explosive growth phase).
Thus, it turns out that the analytic approach is not very
accurate because it underestimates the allowed region
by up to several orders of magnitude in $\Delta m^2_{\tau s}$.
However, it does provide useful qualitative insight, which
is why we included it in Ref.\cite{fv}.

This analytic approach has
recently been re-examined by Shi and Fuller in Ref.\cite{sf}.  
Using a similar set of approximations, {\it including}
Eq.(\ref{x1}), they find the 
slightly less stringent bound
\begin{equation}
|\Delta m^2_{\tau s}/\text{eV}^2| \stackrel{>}{\sim}
3 \times 10^5 \left(|\Delta m^2_{\mu
s}/\text{eV}^2|\right)^{12/11}.
\label{y1}
\end{equation}
Taking $\Delta m^2_{\mu s} = 10^{-3} \ \text{eV}^2$, Eq.(\ref{y1}) then
implies a lower bound on $m_{\nu_{\tau}}$ of about 15 eV. 
However, as we have explained above, and as was also
pointed out in Ref.\cite{fv} (see pages 5167 and 5170), this
analytic approach significantly underestimates the
allowed region because the approximations upon
which it is based are simply not valid. The main
problem, as discussed above, is the assumption
that the $\nu_\tau$ asymmetry is in the $T^{-4}$ 
phase when $\nu_{\mu} \to \nu_s$
oscillations are most strongly acting to produce 
a compensating $L_{\nu_\mu}$ asymmetry.

This point is easy to demonstrate. Consider, by 
way of concrete example, $\nu_{\tau} \to \nu_s$
oscillations with the 
parameters $\Delta m^2_{\tau s} = - 50\ \text{eV}^2$ and $\sin^2
2\theta_{\tau s} = 10^{-8}$. In the 
absence of other oscillation modes involving either
$\nu_{\tau}$ or $\nu_s$, the growth of the $\nu_\tau$
asymmetry is illustrated in Fig.1 of Ref.\cite{f}
which we reproduce here for convenience. In this
example the critical temperature is about 37 MeV, and
the value of the tau-like asymmetry $L_{\nu_{\tau}}$ is about $10^{-5}$ 
when the explosive growth
phase gives way to $T^{-4}$ growth.

Consider now maximally-mixed $\nu_{\mu} \to \nu_s$ oscillations. 
The MSW resonance condition
for this mode is given by
\begin{equation}
V(p) = V_{\text{Wolfenstein}} + V_{\text{finite-T}}
= 0
\label{res}
\end{equation}
where
\begin{equation}
V_{\text{Wolfenstein}} \simeq 
\pm \sqrt{2} G_F n_{\gamma} (2L_{\nu_\mu} + L_{\nu_\tau})
\end{equation}
and
\begin{equation}
V_{\text{finite-T}} 
= -\sqrt{2} G_F n_{\gamma} 
A \left(\frac{T}{m_W}\right)^2 \frac{p}{\langle p
\rangle}
\end{equation}
are the Wolfenstein (finite-density) and finite-temperature 
contributions to the effective matter potential\cite{nr}, 
respectively [the $+$ ($-$) sign refers to neutrino 
(anti-neutrino) oscillations].  The quantity $G_F$ is the 
Fermi constant, $A \simeq 15.3$, $m_W$ is the $W$-boson 
mass, $p$ is the neutrino
momentum, and $\langle p \rangle \simeq 3.15T$ is its thermal 
average. 
Taking for definiteness that $L_{\nu_\tau} > 0$, which
means that the MSW resonance for $\nu_\mu \to \nu_s$
oscillations occurs for the neutrinos rather than
the anti-neutrinos, Eq.(\ref{res})
can be solved for the resonance momentum $p_{\text{res}}$ to yield
\begin{equation}
\frac{p_{\text{res}}}{\langle p \rangle} 
\simeq \frac{L_{\nu_{\tau}} m_W^2}{A T^2}
\end{equation}
provided that $L_{\nu_{\mu}} \ll L_{\nu_{\tau}}$. 
Putting in $L_{\nu_{\tau}} \sim 10^{-5}$ and $T \sim 37$ MeV we find that
\begin{equation}
\frac{p_{\text{res}}}{\langle p \rangle} \sim 3\ \Rightarrow\ 
\frac{p_{\text{res}}}{T} \sim 10.
\label{er}
\end{equation}
This means that as $L_{\nu_{\tau}}$ develops from close to $0$ 
to about $10^{-5}$ during the explosive growth
phase, the MSW resonance momentum for $\nu_{\mu} \to \nu_s$ 
oscillation sweeps through most of
the Fermi-Dirac distribution. This is what we mean when 
we say that the tau-like asymmetry is
still in the explosive growth phase 
when $\nu_{\mu} \to \nu_s$ oscillations are most strongly
acting to produce a compensating $L_{\nu_\mu}$ asymmetry.

Of course, the above was derived assuming that $\nu_{\mu} \to \nu_s$ 
oscillation are not strong
enough to create a 
significant compensating $L_{\nu_{\mu}}$, 
so that $L_{\nu_{\mu}} \ll L_{\nu_{\tau}}$ always holds. 
This is of course the main point. Are
$\nu_\mu \to \nu_s$ oscillations effective enough so that
they create significant $L_{\nu_\mu}$ so that
$p_{\text{res}} \stackrel{<}{\sim} \langle p \rangle$
contrary to Eq.(\ref{er})? In order
to work out exactly what happens, a detailed numerical 
calculation must be performed, which is one of the points 
we were emphasising above.  However,
the above argument clearly shows that $L_{\nu_{\tau}}$ 
should not be taken to be in the $T^{-4}$
phase. Note that if $\sin^2 2\theta_{\tau s} > 10^{-8}$
then the explosive growth phase typically continues 
until even larger values of $L_{\nu_\tau}$ are reached
(i.e. larger than $10^{-5}$).
This obviously makes the $T^{-4}$ growth
assumption even less valid.

To summarise, $\nu_\mu \to \nu_s$ oscillations can solve the
atmosheric neutrino problem without leading to $N_{\text{eff}} = 4$
neutrinos in the early Universe. This result occurs because
$\nu_\tau \to \nu_s$ oscillations generate a large
$L_{\nu_\tau}$ asymmetry (provided
$m_{\nu_\tau} > m_{\nu_s}$). The large $L_{\nu_\tau}$ asymmetry
will suppress the $\nu_\mu \to \nu_s$ oscillations in
the early Universe provided that the $\nu_\mu \to \nu_s$
oscillations do not generate a large $L_{\nu_\mu}$ which
compensates for the effect of the large $L_{\nu_\tau}$ 
in the matter term for $\nu_\mu \to \nu_s$ oscillations.
Calculating the `allowed region' is best handled numerically
and has already been done in Refs.\cite{fv,f}. 
The analytic approach of Shi and Fuller \cite{sf}, which
is based on our own analytic work, significantly underestimates
the allowed region because it uses simplifying assumptions
which are not valid.
We conclude that {\it there is at present no 
cosmological objection to the
$\nu_{\mu} \to \nu_s$ solution to the atmospheric neutrino problem.}

\newpage
{\bf Figure caption}
\vskip 0.5cm
\noindent
Figure 1: The evolution of $|L_{\nu_\tau}|$ for the 
example, $\Delta m^2_{\tau s} = -50 \ eV^2$, $\sin^2 2\theta_{\tau s} =
10^{-8}$.

\newpage
\epsfig{file=f1x.eps, width=15cm}

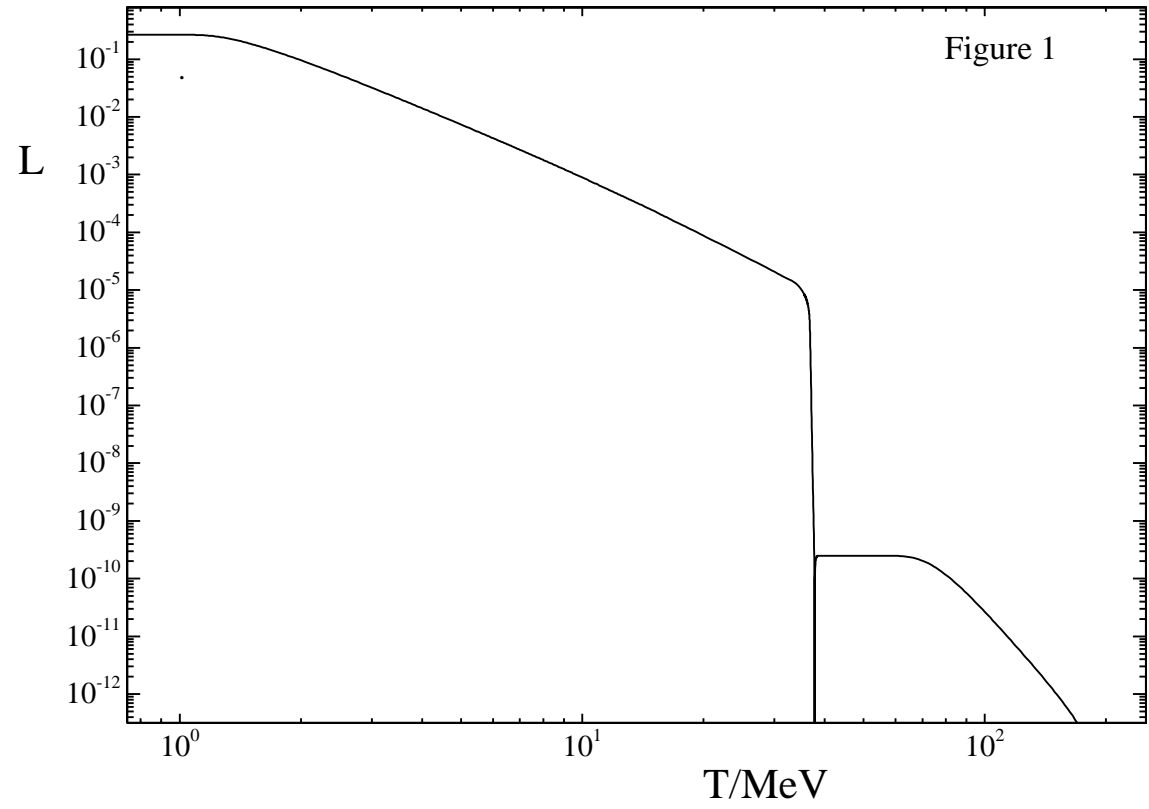
\end{document}